\documentclass[sigconf]{acmart}
\AtBeginDocument{%
  }

\copyrightyear{2026}
\acmYear{2026}
\setcopyright{cc}
\setcctype{by}
\acmConference[KDD '26]{Proceedings of the 32nd ACM SIGKDD Conference on Knowledge Discovery and Data Mining V.2}{August 09--13, 2026}{Jeju Island, Republic of Korea}
\acmBooktitle{Proceedings of the 32nd ACM SIGKDD Conference on Knowledge Discovery and Data Mining V.2 (KDD '26), August 09--13, 2026, Jeju Island, Republic of Korea}
\acmDOI{10.1145/3770855.3817892}
\acmISBN{979-8-4007-2259-2/2026/08}

\usepackage{multirow}
\usepackage{tabularx}
\usepackage{enumitem}
\usepackage{booktabs}
\usepackage{xparse}
\usepackage{graphicx}
\usepackage{subcaption}

\usepackage{listings}
\usepackage{xcolor}

\usepackage{enumitem}

\begin{document}

\title{CoPersona: Collaborative Persona Graphs \\ for Robust LLM Personalization}


\author{Yangtian Zhang}
\authornote{Both authors contributed equally to this research.}
\email{yangtian.zhang@yale.edu}
\orcid{0000-0003-4969-6670}
\affiliation{%
  \institution{Yale University}
  \city{New Haven}
  \state{Connecticut}
  \country{USA}
}

\author{Leyao Wang}
\authornotemark[1]
\email{leyao.wang.lw855@yale.edu}
\affiliation{%
  \institution{Yale University}
  \city{New Haven}
  \state{Connecticut}
  \country{USA}
}

\author{Hiren Madhu}
\email{hiren.madhu@yale.edu}
\affiliation{%
  \institution{Yale University}
  \city{New Haven}
  \state{Connecticut}
  \country{USA}
}

\author{Ngoc Bui}
\email{ngocbh.pt@gmail.com}
\affiliation{%
  \institution{Yale University}
  \city{New Haven}
  \state{Connecticut}
  \country{USA}
}

\author{Walter Roznyatovskiy}
\email{vlad.r@samsung.com}
\affiliation{%
  \institution{Samsung}
  \city{Mountain View}
  \state{California}
  \country{USA}
}

\author{Rex Ying}
\email{rex.ying@yale.edu}
\affiliation{%
  \institution{Yale University}
  \city{New Haven}
  \state{Connecticut}
  \country{USA}
}

\renewcommand{\shortauthors}{Zhang et al.}

\newcommand{\yangtian}[1]{{\color{purple}yangtian: #1}}
\newcommand{\laura}[1]{{\color{teal}Laura: #1}}
\newcommand{\ngoc}[1]{{\color{blue}(Ngoc: #1)}}
\newcommand{\ours}{CoPersona }

\begin{abstract}

Real-world LLM personalization is often constrained by sparse and skewed user histories: most users provide only a handful of interactions, while even frequent users’ logs capture an incomplete and biased view of their preferences. As a result, weakly observed user attributes are difficult to infer, leading to brittle personalization when test-time requests shift toward under-supported facets.

Motivated by this limitation, we present \textbf{CoPersona}, a graph-based collaborative personalization framework that completes sparse user profiles by borrowing signals from behaviorally similar peers. However, directly transferring signals is difficult because uneven facet coverage introduces bias into interaction histories, obscuring user similarity in the unstructured global space. To address this issue, CoPersona decomposes interaction histories into multiple facet-level representations and explicitly models peer-to-peer, facet-level alignment through a multiplex persona graph. To effectively leverage peer information at inference time, we employ a dual-branch architecture that combines \textit{non-parametric peer retrieval} with \textit{parametric graph reasoning}. Experiments across multiple domains and model scales demonstrate consistent improvements over strong baselines, validating CoPersona as an effective approach for robust LLM personalization.

\end{abstract}




\maketitle

\section{Introduction}
Large language models (LLMs) have rapidly expanded in capability, powering applications from conversational assistants and writing tools to decision support and recommendation. Despite these advances, most LLMs remain largely \textit{preference-agnostic}: they are optimized for population-level behavior and often overlook the goals, values, and writing style of a particular individual~\citep{chen2024large, santurkar2023whose, bui2025mixture}. In many practical settings, however, the same request can warrant different responses for different users, motivating \emph{LLM personalization}—adapting generation using user-specific information to improve usefulness and user-level alignment~\citep{aher2023using, lyu2024llm, yang2023palr}.

A dominant line of work~\citep{lamp, longlamp, pag, li2023teach, mysore2024pearl} instantiates personalization through a memory--retrieval pipeline: given a user’s current query, the system retrieves relevant snippets from the user’s interaction history and concatenates them with the query as context for the LLM. 
While effective, real-world histories are often limited in volume and scope~\citep{sun2025persona, shi2025retrieval} and therefore provide a partial and biased view of a user’s persona. A persona spans multiple behavioral dimensions (e.g., preferences, tone, values), but histories typically over-represent some dimensions while under-representing others (e.g., abundant signals about entertainment preferences but little about value orientation). This creates a \emph{facet-level cold-start} regime~\citep{schein2002methods}
For instance, a user's history might clearly indicate a preference for Sci-Fi (Genre Facet) but contain no evidence regarding their critical writing style (Tone Facet). When a query hinges on these missing dimensions, standard retrieval fails, forcing the LLM to revert to generic generation or hallucinate a persona that contradicts the user’s actual voice.

Motivated by collaborative filtering in recommendation systems~\citep{bobadilla2012collaborative}, we investigate a collaborative remedy to this limitation: augmenting a target user’s profile by borrowing information from other users with similar behaviors. Recent work~\citep{dpl, sun2025persona, shi2025retrieval} has begun to incorporate inter-user signals for LLM personalization. However, most existing approaches~\citep{sun2025persona, shi2025retrieval} still depend on shallow retrieval of peers’ documents under a largely global notion of similarity, offering limited structure or interpretability about \textit{which} facets drive similarity and \textit{what} signals should be transferred. This becomes especially problematic under sparse and skewed histories: users may appear similar overall while differing sharply by facet simply because their histories emphasize different aspects (e.g., one reveals more about communication style while another reveals more about preferences). As a result, unstructured neighbor borrowing can be noisy, difficult to control, and hard to audit.

\begin{figure}[t]
  \centering
  \includegraphics[width=\columnwidth]{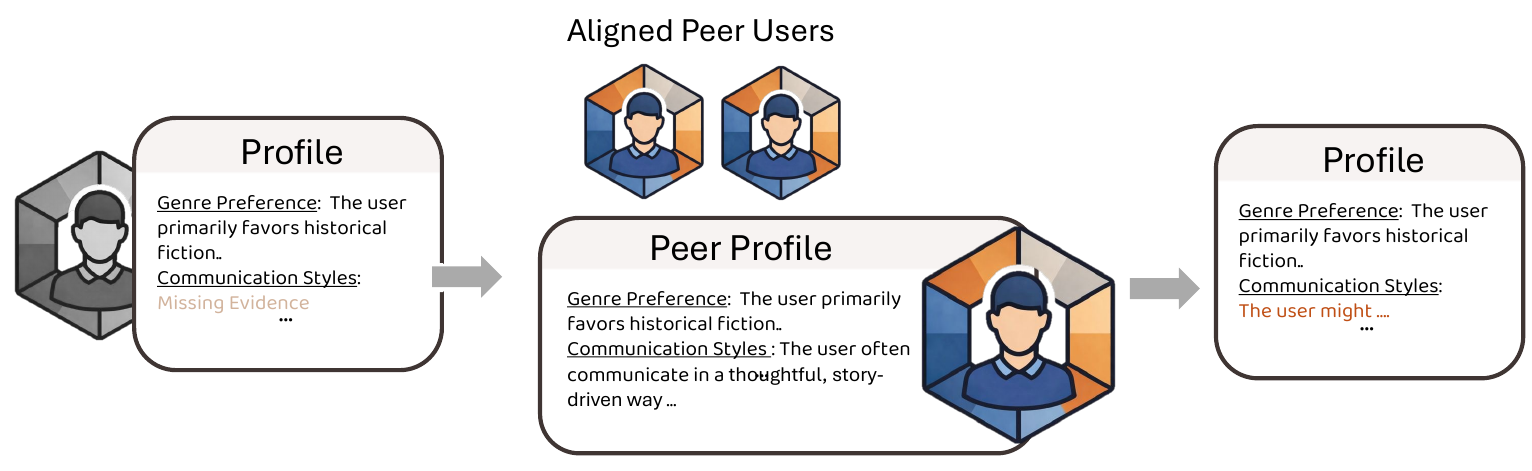}
  \caption{
  Facet coverage bias and collaborative persona completion. By leveraging behaviorally aligned peers, complementary facet evidence can be inferred
  }
  \label{fig:facet_bias_intro}
  \vspace{-10pt}
\end{figure}

To address this challenge, we present \textsc{CoPersona}, a collaborative personalization framework that incorporates persona facet structure via a facet-level similarity graph. \textsc{CoPersona} first discovers a small set of human-interpretable facets from population data, capturing key dimensions along which user personas vary. It then summarizes each user’s history into facet-specific profiles and embeds them to build a multiplex user--user graph, where each layer encodes similarity under a particular facet. This design provides a more interpretable and controllable mechanism for peer borrowing: it enables facet-aligned transfer, reduces negative transfer when users are similar in some dimensions but not others, and remains robust when a user’s own history is sparse or skewed toward a subset of facets.

To leverage peer information effectively at inference time, \textsc{CoPersona} combines two complementary components. (i) A lightweight \emph{non-\allowbreak parametric} branch retrieves facet-aligned neighbor profiles to provide transparent, human-readable evidence that can be directly inspected in the prompt. Using prompt-based evidence alone can result in performance limitation under context-length and noise constraints: relevant peer signals may be sparse, redundant, or difficult for the LLM to integrate consistently. (ii) A complementary \emph{parametric} graph branch performs reliability-gated message passing over the multiplex facet relationship graph, aggregating neighbor information in latent space to complement under-supported facets and compressing the result into soft-prompt tokens. Together, these components mitigate facet coverage bias and improve personalization when user histories are sparse or uneven. Experiments across multiple domains and model scales show consistent gains over strong baselines, validating \textsc{CoPersona} as a practical approach to robust LLM personalization via \emph{graph-based peer borrowing}. In summary, our contributions are:

\begin{itemize}[leftmargin=*]\setlength\itemsep{0.15em}
    \item We propose \textsc{CoPersona}, which discovers persona facets, constructs a facet-level multiplex similarity graph, and enables inference-time graph-based collaborative information borrowing.
    \item We develop a dual-branch inference mechanism—(i) facet-aligned neighbor retrieval in text space and (ii) reliability-gated graph aggregation in latent space—that robustly complements missing or weakly observed facets.
    \item We demonstrate consistent gains across domains and model scales, validating \textsc{CoPersona} as a practical approach to robust LLM personalization.
\end{itemize}

\section{Preliminaries}
\subsection{Problem Formulation}
\label{sec:problem_formulation}

LLM personalization aims to generate user-aligned text by leveraging a user’s historical interactions.
Let $\mathcal{U}=\{u_1,\ldots,u_N\}$ denote users. For each user $u$, we observe a history
$\mathcal{D}_u=\{(x_k,y_k)\}_{k=1}^{N_u}$, where each pair corresponds to one interaction.
Here $x$ is the input context that specifies \emph{what to respond to} (e.g.,
an item title/category/short description for review generation, or an incoming email/dialogue history for assistant-style tasks).
The target $y$ is the user-authored text response (e.g., a critical vs.\ enthusiastic review, a terse vs.\ detailed reply),
which implicitly reflects the user’s preferences and writing style.

Standard methods assume that the observed history $\mathcal{D}_u$ is a sufficient statistic to model the user's behavior. 
They learn parameters $\theta$ of a conditional model $P_\theta$ by maximizing the likelihood of observed responses given queries and the user's history:
\begin{equation}
\theta^\star=\underset{\theta}{\operatorname{argmax}} \sum_{(x,y)\in \mathcal{T}_u} \log P_\theta\left(y \mid x, \mathcal{D}_u\right),
\end{equation}
where $\mathcal{T}_u$ denotes training examples for user $u$ (or a population-level training set with user histories).

\subsection{Facet Coverage Bias}
\label{sec:facet_bias}
In reality, user personas are inherently multi-faceted rather than monolithic: a single user may vary along multiple latent behavioral dimensions (e.g., topical preferences, value orientation, tone, and stylistic habits), and different generation contexts can selectively elicit different dimensions. Following \citet{lu-etal-2023-miracle, chen2024recenttrendspersonalizeddialogue}, we view each user as governed by a set of latent facets that jointly shape their outputs.

Concretely, let $\mathcal{F}=\{f_m\}_{m=1}^{M}$ denote a \emph{global} facet schema. For each user $u$, we associate a latent persona state $Z_u=\{z_u^{(m)}\}_{m=1}^{M}$, where $z_u^{(m)}$ captures how user $u$ behaves along facet $f_m$. We observe only a finite interaction history $D_u=\{(x_k,y_k)\}_{k=1}^{N_u}$, where each interaction provides \emph{partial} and context-dependent evidence about $Z_u$. Importantly, the history is rarely uniformly informative across facets~\citep{chen2021biasdebiasrecommendersystem, 10.1145/3336191.3371783}: some facets are repeatedly surfaced by the user’s logged contexts (e.g., consistent formality and analytical style in product reviews), while other facets (e.g., humor usage, emotional expressiveness, or conversational warmth) may be sparsely exhibited or never elicited.

We refer to this uneven observability as \textbf{Facet Coverage Bias}: the empirical distribution of evidence in $D_u$ is skewed toward a subset of facets, yielding low coverage for the remainder. Low coverage should be interpreted as \emph{under-observation rather than absence}~\citep{chen2021biasdebiasrecommendersystem, 10.1145/3336191.3371783}—a facet can be part of the user’s true persona yet remain weakly supported in the recorded log. This perspective is consistent with prior findings that observational user traces are selective and incomplete, where non-observation does not imply non-preference or non-existence.

Facet coverage bias becomes problematic at test time when the new query $x'$ implicitly requires facets that are weakly supported in $D_u$. In this regime, standard personalization methods (retrieval-, profile-, or embedding-based) tend to extrapolate from the dominant observed facets, causing \emph{over-transfer}: the model produces outputs that are faithful to what is well-covered in the history (e.g., persistent formality) but misaligned with the facet(s) that the current request actually hinges on (e.g., a more playful, empathetic, or persuasive tone). In Sec.~\ref{sec:facet_coverage_analysis}, we empirically quantify this coverage skew and show that performance degradation concentrates on under-covered facets, motivating our facet-aware collaborative completion strategy.

\section{Methodology}

\begin{figure*}[t]
  \centering
  \includegraphics[width=0.95\textwidth]{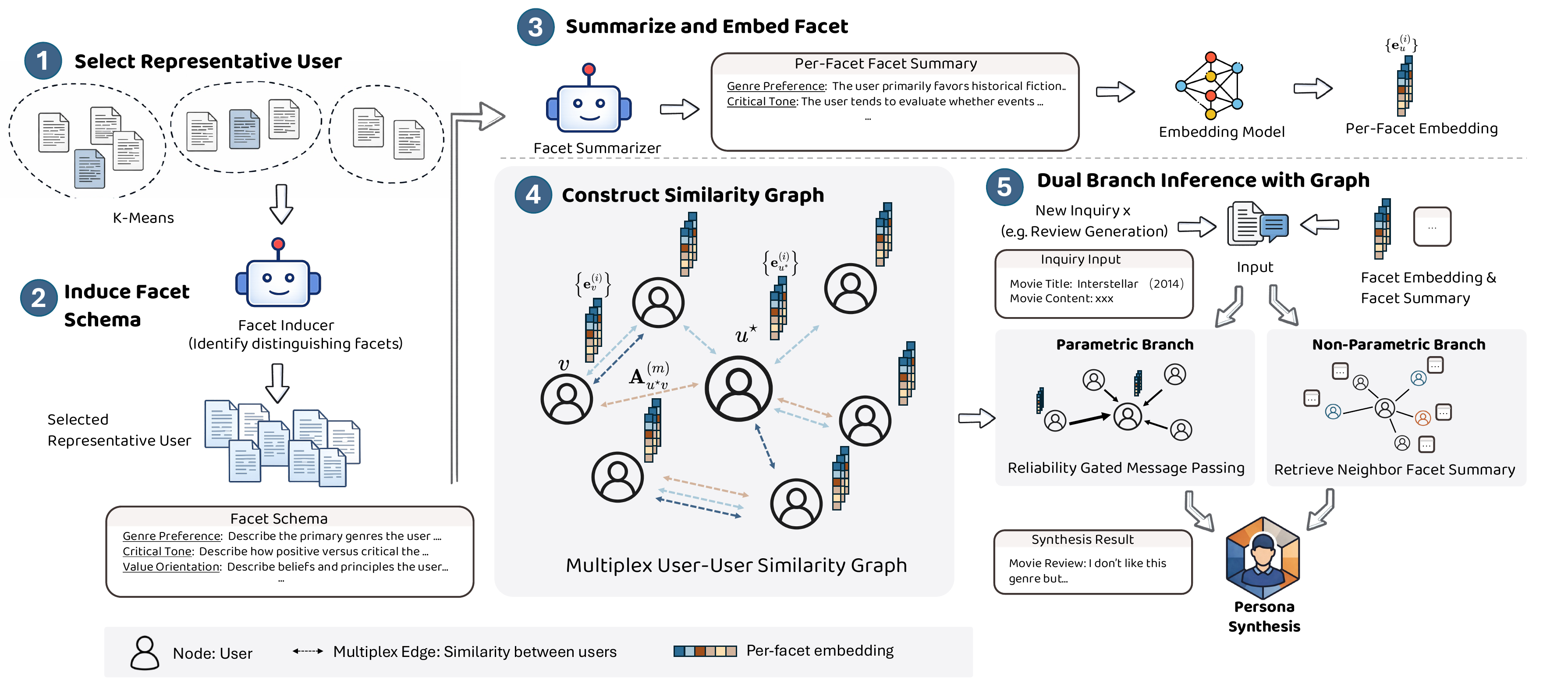}
  \vspace{-5pt}
  \caption{CoPersona overview. CoPersona personalizes an LLM by (1) mining representative users and inducing a global facet schema, (2) summarizing each user per facet and constructing a multiplex user–user similarity graph (one layer per facet), and (3) performing dual-branch inference that combines non-parametric facet-level neighbor retrieval with parametric graph reasoning (e.g., reliability-gated message passing) to synthesize a personalized persona signal for generation. }
  \label{fig:copersona}
\end{figure*}

\label{sec:method_overview}
To mitigate \textit{Facet Coverage Bias} arising from isolated and often sparse user histories, we propose \textbf{CoPersona}, a framework that casts LLM personalization as \emph{graph-based collaborative inference}. The core insight is that, although any single user’s history may be limited, population-level behavior is comparatively dense and structured: users who align on \emph{observed} facets (e.g., value orientation) are more likely to share \emph{unobserved} facets (e.g., genre preference). Accordingly, instead of restricting personalization for user $u$ to their own history $\mathcal{D}_u$, CoPersona models the population manifold as a \textbf{Collaborative Persona Graph} $\mathcal{G}$---a multiplex user--user similarity graph where each facet induces a similarity layer connecting semantically aligned users. This graph acts as a semantic bridge that enables a target user to “borrow” complementary signals from similar peers to complete their latent persona for downstream generation. Formally, we expand the probabilistic generation model to condition on $\mathcal{G}$:
\begin{equation}
    P_\theta\!\left(y' \mid x', \mathcal{D}_u, \mathcal{G}\right),
\end{equation}
from which a response is produced via a decoding strategy (e.g., sampling or beam search). 
CoPersona instantiates this formulation in three phases: (1) \textit{Facet Disentanglement} (Sec.~\ref{sec:facet_discovery}), which structures raw histories into facet-specific summaries and embeddings; (2) \textit{Multiplex Graph Construction} (Sec.~\ref{sec:persona_graph}), which maps inter-user per-facet relations; and (3) \textit{Dual-Branch Inference} (Sec.~\ref{sec:dual_branch_inference}), which jointly leverages parametric reasoning and non-parametric context retrieval.

\subsection{Facet Disentanglement}
\label{sec:facet_discovery}

User personas are inherently multi-faceted, yet real-world histories are sparse and skewed: only a few behavioral attributes are repeatedly expressed, while others are rarely elicited. CoPersona therefore \emph{disentangles} a user’s history into facet-specific representations in two steps: we first induce a compact \emph{global facet schema} in a data-driven manner, then summarize and embed each user \emph{per facet} to obtain a structured persona state.

\paragraph{Facet schema induction.}
A key design choice is to avoid a manually pre-defined facet list. Predefining facets is brittle: the “right” schema is domain- and population-dependent, hard to enumerate exhaustively, and easy to miss emergent behavioral axes that matter for downstream generation. Instead, we induce facets from population data so the schema (i) adapts to the dataset’s dominant modes of variation, (ii) remains human-interpretable, and (iii) provides consistent organizing dimensions for facet-aware similarity and collaboration.

Concretely, we first construct a profile text $\tau_u$ for each user by concatenating up to $R$ historical interactions. We encode $\tau_u$ with a sentence encoder to obtain a dense profile embedding $z_u \in \mathbb{R}^d$, and cluster $\{z_u\}$ into $K$ groups using standard clustering algorithm (e.g., KMeans or HDBSCAN) to identify representative persona modes. 

For each cluster $k$, we select a small set of prototype users and build a compact evidence package $S_k = (T_k, E_k)$ where $T_k$ is a set of cluster-level keywords/phrases summarizing salient cues (i.e. top TF-IDF terms computed over $\{\tau_u : u \in \text{Cluster }k\}$),
and ${E}_k$ is a small set of representative evidence (e.g., short excerpts) collected from users in cluster $k$.

Collecting these packages as $P=\{S_k\}_{k=1}^{K}$, we prompt an LLM to \emph{contrast} clusters and induce a facet schema whose axes best separate the representative persona modes:
\begin{equation}
    \hat{\mathcal{F}}=\{f_m\}_{m=1}^{M} = \operatorname{LLM}{({P})},
\end{equation}
where each facet $f_m$ includes (i) a concise name, (ii) a natural-language definition describing the behavioral attribute, and (iii) a scoring specification (categorical or ordinal) to clarify how evidence should be interpreted. We further instruct the LLM to avoid redundant facets and prefer orthogonal dimensions (e.g., tone vs.\ topical preference vs.\ value orientation), yielding a small, interpretable schema that is shared across users and serves as the organizing backbone for downstream per-facet summarization and facet-wise similarity modeling.

\paragraph{Facet summarization, reliability, and embeddings.}
Given the induced schema $\mathcal{F}$, we derive facet-specific summarization and representations for every user. For each user $u$ and facet $m$, we gather the set of historical interactions and prompt an LLM to produce: (i) a concise facet summary $s_u^{(m)}$ describing the user’s typical behavior along the facet dimension $f_m$ (e.g., value orientation), (ii) optionally up to two short supporting quotes $\mathrm{ev}_u^{(m)}$ grounded in the user’s past text, and (iii) a reliability label $r_u^{(m)} \in \{\texttt{none}, \texttt{weak}, \texttt{moderate}, \texttt{strong}\}$ that indicates how well the facet summary is supported by available evidence. We form the facet description by concatenation:
\begin{equation}
    t_u^{(m)} = [\, s_u^{(m)} ; \mathrm{ev}_u^{(m)} \,].
\end{equation}
We then embed $t_u^{(m)}$ with a pretrained sentence encoder and $\ell_2$ normalization to obtain a dense per-facet embedding $e_u^{(m)} \in \mathbb{R}^d$. For facets with little or no usable evidence (typically $r_u^{(m)}\in\{\texttt{none},\texttt{weak}\}$), we instantiate a neutral placeholder description (e.g., ``No explicit profile information provided.'') before encoding, ensuring every user has a well-formed representation on every facet. The resulting embeddings $\{e_u^{(m)}\}_{m=1}^M$, together with reliability labels $\{r_u^{(m)}\}_{m=1}^M$, constitute a disentangled persona state; in later stages, the reliability signal is used to downweight or filter low-support facets when constructing facet-wise neighborhoods and performing collaborative inference.

\subsection{Multiplex Collaborative Persona Graph}
\label{sec:persona_graph}

Given the facet-level decomposition from Sec.~\ref{sec:facet_discovery}, we construct a \emph{multiplex} Collaborative Persona Graph $\mathcal{G}$ that organizes users by their similarity \emph{within each facet}. Concretely, $\mathcal{G}$ contains $M$ facet-induced layers, one per facet $m\in\{1,\dots,M\}$ in the induced schema $\mathcal{F}$. Each layer shares the same node set $\mathcal{V}$ (one node per user), but has its own edge set capturing facet-specific peer alignment. This multiplex structure provides the backbone for controlled, facet-aware collaboration: it allows CoPersona to borrow complementary signals from peers on the particular facet that is under-observed for the target user, rather than collapsing all evidence into a single global similarity score.

\paragraph{Nodes and facet states.}
For every user $u$ and facet $m$, Sec.~\ref{sec:facet_discovery} produces a normalized facet embedding $\mathbf{e}_u^{(m)}\in\mathbb{R}^d$ together with a reliability label
$r_u^{(m)} \in \{\texttt{none},\texttt{weak},\texttt{moderate},\texttt{strong}\}$ indicating how well the facet summary is supported by the user history. We use $\{\mathbf{e}_u^{(m)}\}_{u\in\mathcal{V}}$ as the node embedding table for layer $m$, and leverage $\{r_u^{(m)}\}$ to downweight or filter low-evidence facets when forming neighborhoods.

\paragraph{Facet-wise edges and sparsification.}
Edges are defined independently within each facet layer. In layer $m$, we compute the cosine similarity between users $u$ and $v$ using their normalized facet embeddings:
\begin{equation}
\label{eq:sim_1}
\mathbf{A}_{uv}^{(m)} = \text{sim}(\mathbf{e}_u^{(m)}, \mathbf{e}_v^{(m)}) = \frac{\mathbf{e}_u^{(m)} \cdot \mathbf{e}_v^{(m)}}{\|\mathbf{e}_u^{(m)}\| \|\mathbf{e}_v^{(m)}\|}.
\end{equation}
To incorporate evidence quality, we optionally apply a reliability gate to the similarity by multiplying by a scalar weight $\omega(\cdot)$ (Details in Appendix.~\ref{apd:implementation}):
\begin{equation}
\label{eq:sim_2}
\tilde{\mathbf{A}}_{uv}^{(m)} \;=\; \omega\!\left(r_u^{(m)}\right)\,\omega\!\left(r_v^{(m)}\right)\,\mathbf{A}_{uv}^{(m)},
\end{equation}
where $\omega(\texttt{none}) \le \omega(\texttt{weak}) \le \omega(\texttt{moderate}) \le \omega(\texttt{strong})$.
Finally, to reduce noise and ensure scalability, we sparsify each layer by retaining only the top-$K$ neighbors per user under $\tilde{\mathbf{A}}^{(m)}$, yielding facet-specific neighborhoods $\mathcal{N}_u^{(m)}$ and a sparse adjacency per layer.

Under this multiplex construction, when a target user under-expresses a facet $m$ (low $r_u^{(m)}$), CoPersona can still identify a set of aligned peers using \emph{other} facets $m' \neq m$ that are well-supported for the user (high $r_u^{(m')}$). This allows CoPersona to implicitly infer facet $m$ by aggregating these peers' facet-$m$ representations.

\subsection{Dual-Branch Inference}
\label{sec:dual_branch_inference}
CoPersona combines two complementary inference mechanisms. The non-parametric branch retrieves facet-specific exemplars from behaviorally similar users, preserving human-interpretable signals in observed facets. The parametric branch performs latent collaborative inference on the Collaborative Persona Graph, using reliability-gated message passing to denoise observed facets and impute missing ones in a continuous embedding space. At inference time, both branches contribute conditioning signals to the base LLM: retrieved exemplars are provided as textual context, while the collaborative persona embedding is injected as soft prompts.

\paragraph{Non-Parametric Facet Retrieval}
\label{sec:nonparametric_branch}

The first branch of CoPersona implements a non-parametric, retrieval-style form of personalization that operates directly in the text space by reusing the facet-wise neighborhoods of the multiplex Collaborative Persona Graph (Sec.~\ref{sec:persona_graph}). In facet layer $m$, edges are weighted by the reliability-gated facet similarity $\tilde{A}_{u v}^{(m)}$ derived from cosine similarity of facet embeddings (Eq.~\ref{eq:sim_1} and Eq.~\ref{eq:sim_2}).

For a target user $u^{\star}$ and facet $m$, we retrieve neighbors directly from the sparsified graph:
\begin{equation}
    {N}_{u^{\star}}^{(m)}=\left\{v \mid\left(u^{\star}, v\right) \in \mathcal{E}^{(m)}\right\},
\end{equation}
where $\mathcal{E}^{(m)}$ retains the top- $K$ neighbors per user under $\tilde{A}^{(m)}$. (Equivalently, ${N}_{u^{\star}}^{(m)}$ matches the TopK set under $\tilde{A}_{u^{\star} v}^{(m)}$; we use the stored neighborhood for efficiency and consistency.)

From each neighbor $v \in N^{(m)}_{u^\star}$, we retrieve a facet-aligned
\emph{exemplar package} $t^{(m)}_v = [s^{(m)}_v; ev^{(m)}_v]$ (Eq.~4), which
includes the neighbor’s facet summary and (when available) short evidence
quotes grounded in their past text.

We then form the facet-wise exemplar set:
\begin{equation}
E^{(m)}_{u^\star} =
\left\{ \bigl(v,\; t^{(m)}_v,\; \tilde{A}^{(m)}_{u^\star v}\bigr)\;\middle|\; v \in N^{(m)}_{u^\star} \right\}.
\end{equation}

Finally, we linearize $\{E^{(m)}_{u^\star}\}_{m=1}^M$ into a facet-grouped peer
context $C^{\text{peer}}_{u^\star}$, where $\tilde{A}^{(m)}_{u^\star v}$ serves as an
explicit confidence signal over retrieved peer exemplars. This branch makes no parametric updates and preserves facet-aligned, human-interpretable peer signals.

\paragraph{Parametric Graph Reasoning}
\label{sec:parametric_graph}
While the non-parametric branch can already supply useful signals for under-observed facets by retrieving facet-specific exemplars from similar users, it does so in an explicitly local and token-level manner, and its effectiveness is constrained by retrieval noise and context length. The parametric branch of CoPersona is designed as a complementary mechanism: instead of operating on raw text, it performs latent collaborative inference over the Collaborative Persona Graph via a \textbf{reliability-gated message passing scheme}, aggregating information from neighbors into continuous facet representations and learning how strongly to trust self versus peers. This yields smoothed, reliability-aware facet embeddings that provide a compact soft-prompt prior, complementing the discrete, example-based guidance from the non-parametric branch.

We start from the facet-specific embeddings $\mathbf{e}_u^{(m)}$ and the ordinal reliability label $r_u^{(m)} \in$ \{none, weak, moderate, strong\} introduced in Sec.~\ref{sec:facet_discovery}, and use it to control how strongly the parametric branch trusts the user's own facet representation versus information aggregated from neighbors. Concretely, we transform the discrete label into a scalar trust gate $\gamma_u^{(m)}=f_\theta\left(r_u^{(m)}\right) \in[0,1]$ using a learnable monotone scoring function $f_\theta(\cdot)$\footnote{Constrained to satisfy $f_\theta($ none $) \leq f_\theta($ weak $) \leq f_\theta($ moderate $) \leq f_\theta($ strong $)$. See details in Appendix.~\ref{apd:implementation}}, . This yields a monotonically increasing confidence score aligned with evidence strength, and $\gamma_u^{(m)}$ is then used in the reliability-gated message passing update to interpolate between the user's own facet embedding and the neighbor-aggregated embedding, with larger $\gamma_u^{(m)}$ placing more weight on self and smaller values relying more on peers.

To exploit cross-user structure, we construct a shared neighbor set $\mathcal{N}_u$ at the user level (e.g., from an anchor embedding that aggregates the observed facets of $u$). Intuitively, $\mathcal{N}_u$ contains users whose overall persona profiles are structurally similar, even if individual facets may be missing for $u$. For each facet $m$, we initialize the node state with the facet embedding,
\begin{equation}
    \mathbf{h}_u^{(0,m)} = \mathbf{e}_u^{(m)},
\end{equation}
and aggregate facet-$m$ information from the neighbors of $u$:
\begin{equation}
    \tilde{\mathbf{h}}_u^{(m)} = \mathrm{Agg}_{v \in \mathcal{N}_u} \mathbf{h}_v^{(0,m)},
\end{equation}
where $\mathrm{Agg}$ is a permutation-invariant operator (e.g., mean). We then apply a reliability-aware, self-vs-neighbors gate to obtain the updated facet representation:
\begin{equation}
    \mathbf{h}_u^{(m)} 
= \sigma\!\Big(
W \big(
\gamma_u^{(m)} \,\mathbf{h}_u^{(0,m)} 
+ (1 - \gamma_u^{(m)}) \,\tilde{\mathbf{h}}_u^{(m)}
\big)
\Big),
\end{equation}
with $W$ a shared projection and $\sigma$ a nonlinearity. When facet $m$ is strongly supported for user $u$ ($\gamma_u^{(m)}\approx 1$), the update is dominated by the user’s own facet embedding and only lightly influenced by neighbors. When the LLM indicates that facet $m$ is weak or absent for $u$ ($\gamma_u^{(m)}\approx 0$), the update leans toward the aggregated neighbor embedding, effectively imputing the missing facet from structurally similar peers.

The resulting set $\{\mathbf{h}_u^{(m)}\}_{m=1}^M$ forms a latent, collaboratively smoothed view of the user’s persona. To capture semantic dependencies between these behavioral dimensions, we apply a cross-facet self-attention mechanism over the facet embeddings after each layer. We finally concatenate these contextualized embeddings to obtain a single collaborative persona vector $\mathbf{h}_u$:
\begin{equation}
    \mathbf{h}_u = \big[ {\mathbf{h}}_u^{(1)} \mathbin{\|} \dots \mathbin{\|} {\mathbf{h}}_u^{(M)} \big].
\end{equation}
This vector is then projected into a fixed number of soft prompt tokens that condition the base LLM during generation. In contrast to the non-parametric branch, which operates in the space of retrieved exemplars, the parametric branch performs inference in a latent space over the graph, denoising observed facets and inferring plausible values for under-observed ones.

\subsection{Joint Optimization}
\label{sec:joint_opt}

During training, we optimize all learnable components of CoPersona jointly with the base LLM. For each user $u$ and query $x'$, the parametric branch produces a collaborative persona vector $\mathbf{h}_u$ from the reliability-gated graph reasoning in Sec.~\ref{sec:parametric_graph}. We map this vector into a fixed-length sequence of soft prompt tokens by a learned linear projection:
\begin{equation}
    \mathbf{S}_u = \mathrm{reshape}\big(W_p \mathbf{h}_u\big) \in \mathbb{R}^{L_p \times d_{\text{LLM}}},
\end{equation}
where $W_p \in \mathbb{R}^{(L_p \cdot d_{\text{LLM}}) \times d_{\text{persona}}}$ is a trainable matrix, $L_p$ is the number of soft prompt tokens, $d_{\text{LLM}}$ is the hidden dimension of the base LLM, and $\mathrm{reshape}(\cdot)$ converts the projected vector into a matrix of $L_p$ token embeddings.

In parallel, we retrieve query-relevant interactions from the target user’s own
history via BM25~\citet{robertson2009probabilistic}, following~\citet{dpl}. Let
$\mathcal{H}_u(x') = \textsc{BM25TopK}(x', \mathcal{D}_u)$ denote the retrieved self-history.
We combine this with the peer exemplar context from the non-parametric branch
to form the textual conditioning:
$
C_u(x') = \textsc{Linearize}\big(\mathcal{H}_u(x'), \{E^{(m)}_u\}_{m=1}^M \big).
$

The final input to the LLM is the concatenation, in embedding space, of the soft
prompts $\mathbf{S}_u$, the retrieved context $C_u(x')$, and the current query $x'$.
We train by minimizing the negative log-likelihood of $y$:
\begin{equation}
\mathcal{L} = - \sum_{t=1}^{|y|}
\log P_{\mathrm{LLM}}\!\left(y_t \mid y_{<t},\, \mathbf{S}_u,\, C_u(x'),\, x'\right).
\end{equation}

\section{Experimental Results}
\begin{table*}[t]
\centering
\small
\setlength{\tabcolsep}{4pt}
\caption{Performance comparison across datasets and model scales.
R-1 denotes ROUGE-1, MET. denotes METEOR, BL. denotes BLEU, and BS. denotes BERTScore.
Best results within each block are highlighted in bold.}
\vspace{-5pt}
\label{tab:persona_results}
\begin{tabular}{llcccccccccccc}
\toprule
\multirow{2}{*}{\textbf{Model}} & 
\multirow{2}{*}{\textbf{Method}} &
\multicolumn{4}{c}{\textbf{Books}} &
\multicolumn{4}{c}{\textbf{Movies \& TV}} &
\multicolumn{4}{c}{\textbf{CDs \& Vinyl}} \\
\cmidrule(lr){3-6} \cmidrule(lr){7-10} \cmidrule(lr){11-14}
& & R-1 & MET. & BL. & BS. & R-1 & MET. & BL. & BS. & R-1 & MET. & BL. & BS. \\
\midrule

\multirow{4}{*}{\textbf{32B}}
& Non-Perso
& 0.3025 & 0.1949 & 2.6728 & 0.4970
& 0.2608 & 0.1666 & 1.1226 & 0.4702
& 0.2765 & 0.1767 & 1.6597 & 0.4742 \\

& RAG
& 0.3404 & 0.2735 & 6.8178 & \underline{0.5159}
& 0.2983 & 0.2142 & 2.8680 & 0.4822
& 0.3092 & 0.2177 & 3.1588 & 0.4868 \\

& PAG
& 0.3276 & 0.2830 & 6.8920 & 0.5051
& 0.2816 & 0.2130 & 2.7751 & 0.4746
& 0.2971 & 0.2215 & 3.2164 & 0.4787 \\

& DPL
& 0.3392 & \underline{0.3003} & \underline{7.7423} & 0.5156
& 0.2967 & \underline{0.2238} & \underline{3.2965} & \underline{0.4855}
& \underline{0.3119} & \underline{0.2337} & \underline{3.8271} & \underline{0.4910} \\

\midrule

\multirow{6}{*}{\textbf{7B}}
& Non-Perso
& 0.2907 & 0.1735 & 1.9766 & 0.5004
& 0.2469 & 0.1503 & 0.7242 & 0.4713
& 0.2604 & 0.1561 & 1.0997 & 0.4753 \\

& RAG
& 0.3149 & 0.2101 & 3.6874 & 0.5083
& 0.2693 & 0.1701 & 1.3021 & 0.4787
& 0.2796 & 0.1733 & 1.6129 & 0.4824 \\

& PAG
& 0.3136 & 0.2378 & 4.6762 & 0.4992
& 0.2761 & 0.1905 & 1.9360 & 0.4735
& 0.2882 & 0.1979 & 2.4740 & 0.4789 \\

& DPL
& 0.3194 & 0.2459 & 5.6623 & 0.5050
& 0.2845 & 0.1958 & 2.2451 & 0.4795
& 0.2952 & 0.2003 & 2.6943 & 0.4838 \\

& DEP
& {0.3621} & {0.3009} & {12.3350} & {0.5496}
& {0.3061} & {0.2372} & {6.5927} & {0.5086}
& {0.3143} & 0.2309 & {6.3548} & {0.5122} \\

& \textbf{CoPersona (Ours)}
& \textbf{0.3953} & \textbf{0.3349} & \textbf{15.4703} & \textbf{0.5729}
& \textbf{0.3142} & \textbf{0.2401} & \textbf{8.0569} & \textbf{0.5184}
& \textbf{0.3216} & \textbf{0.2438} & \textbf{7.8669} & \textbf{0.5177} \\

\bottomrule
\end{tabular}
\end{table*}

\begin{table*}[t]
\centering
\small
\setlength{\tabcolsep}{4pt}
\caption{Performance comparison across datasets and model scales.
R-1 denotes ROUGE-1, R-L denotes ROUGE-L, MET. denotes METEOR, and BL. denotes BLEU.
Best results within each block are highlighted in bold.}
\label{tab:persona_results_new}
\begin{tabular}{llcccccccccccc}
\toprule
\multirow{2}{*}{\textbf{Model}} & 
\multirow{2}{*}{\textbf{Method}} &
\multicolumn{4}{c}{\textbf{Video Games}} &
\multicolumn{4}{c}{\textbf{Musical Instruments}} &
\multicolumn{4}{c}{\textbf{Sports\& Outdoors}} \\
\cmidrule(lr){3-6} \cmidrule(lr){7-10} \cmidrule(lr){11-14}
& & R-1 & R-L& MET.& BL.& R-1 & R-L& MET.& BL.& R-1 & R-L& MET.& BL.\\
\midrule

\multirow{7}{*}{\textbf{7B}}
& Non-Perso
& 0.2386 & 0.1304 & 0.1664 & 0.74
& 0.2388 & 0.1348 & 0.2062 & 1.72
& 0.2752 & 0.1497 & 0.2114 & 1.93 \\

& RAG
& 0.2553 & 0.1365 & 0.1841 & 1.30
& 0.2449 & 0.1369 & 0.2157 & 2.05
& 0.2826 & 0.1528 & 0.2225 & 2.45 \\

& PAG
& 0.2537 & 0.1307 & 0.2008 & 1.99
& 0.2265 & 0.1253 & \textbf{0.2224} & 1.65
& 0.2671 & 0.1427 & \textbf{0.2336} & 1.91 \\

& DPL
& 0.2626 & 0.1366 & 0.2005 & 2.01
& 0.2401 & 0.1353 & 0.2145 & 2.77
& 0.2775 & 0.1507 & 0.2234 & 2.19 \\

& DEP
& 0.2655 & 0.1440 & 0.1974 & 4.66
& 0.2319 & 0.1321 & 0.1977 & 3.07 
& 0.2839 & 0.1738 & 0.2041 & 4.68\\

& \textbf{CoPersona (Ours)}
& \textbf{0.2928} & \textbf{0.1553} & \textbf{0.2086} & \textbf{5.87}
& \textbf{0.2447} & \textbf{0.1503} & 0.1760 & \textbf{5.06}
& \textbf{0.3004}  & \textbf{0.1747} & 0.2140 & \textbf{5.86} \\

\bottomrule
\end{tabular}
\end{table*}

We conduct experiments on real-world personalization benchmarks to answer the following research questions:
\begin{itemize}[leftmargin=*, topsep=2pt, itemsep=0pt]
    \item \textbf{RQ1}: How does \ours compare with existing personalization baselines in overall generation quality and user alignment?
 
    \item \textbf{RQ2}: Do the reviews generated by \ours appear more human-like and authentic, as assessed by LLM-as-Judge evaluation?

   \item \textbf{RQ3}: How much does each component of \ours contribute to performance (e.g., the parametric branch, the non-parametric branch, and the retrieval strategy)?

    \item \textbf{RQ4}: How robust is \ours to key design choices (e.g., the number of facets and neighbors), and how sensitive is it to major hyperparameters (e.g., decoding temperature)?
\end{itemize}

\subsection{Experimental Setup}
\label{sec:exp_setup}
\noindent
\textbf{Datasets.} 
Building upon prior work, we focus on the representative task of review generation for LLM personalization~\cite{dep,dpl,longlamp,amazon2023, ni2019justifying, reviewllm, au2025personalized}. Specifically, we adopt the Amazon Reviews 2023 dataset\footnote{\url{https://amazon-reviews-2023.github.io/}}~\cite{amazon2023}
, as used by DEP~\cite{dep}, covering three categories: \textit{Books}, \textit{Movies \& TV}, and \textit{CDs \& Vinyl}.
In addition, we construct three more category-specific datasets from Amazon Reviews 2023 by applying the DPL preprocessing pipeline~\cite{dpl} to \textit{Video Games}, \textit{Musical Instruments}, and \textit{Sports \& Outdoors}. These categories exhibit sparser user--item interactions, allowing us to better assess generalization under limited history. We intentionally filter the data to ensure that each user retains only a limited number of reviews, thereby maintaining sparsity. For all six datasets, we follow the original test splits provided by DPL~\cite{dpl}. More details about the datasets are provided in Appendix~\ref{apd_dataset}.

\vspace{0.5em}
\noindent
\textbf{Baselines.}
Following existing personalization works such as DPL \cite{dpl} and DEP \cite{dep}, We compare \ours with the following baseline methods (implementation details in Appendix~\ref{apd_baseline}. For some baselines, we also report results using a larger 32B base model.%

\begin{itemize}[leftmargin=*, topsep=4pt, itemsep=0pt]
\item \textbf{Non-Perso}: Generates reviews using only item information, titles, and ratings, without user personalization.
\item \textbf{RAG}~\cite{lamp}: Conditions generation on retrieved user history records for contextual personalization.
\item \textbf{PAG}~\cite{pag}: Summarizes user history into a compact profile and combines it with retrieved evidence for structured personalization.
\item \textbf{DPL}~\cite{dpl}: Builds a textual user profile by contrasting the target user with similar peers and prepends it to the prompt.
\item \textbf{DEP}~\cite{dep}: Learns user embeddings that capture preference differences and injects them into a frozen LLM as soft conditioning signals.
\end{itemize}

\vspace{0.5em}
\noindent
\textbf{Evaluation Metrics.}
Following previous works on personalized text generation~\cite{lamp,longlamp,stylevector, au2025personalized,reviewllm, dpl, dep}, we report ROUGE-1~\cite {rouge}, ROUGE-L~\cite{rouge}, METEOR~\cite{meteor}, BLEU\footnote{We use the standard \texttt{SacreBLEU}~\cite{sacrebleu}.}~\cite{bleu},
and BERTScore\footnote{We adopt \texttt{led-base-16384}~\cite{longformer} to compute embeddings.}~\cite{bertscore}.

In addition, we report \textit{LLM-as-Judge} evaluation to assess the human-likeness of generated reviews beyond n-gram overlap. Unlike BLEU/ROUGE, which capture surface similarity, LLM-as-Judge directly evaluates naturalness and authenticity, providing a scalable proxy for whether reviews sound genuinely human \cite{tseng-etal-2024-two}.

\vspace{0.5em}
\noindent
\textbf{Implementation Details.}
We build facet-specific user profiles by summarizing up to six reviews with GPT-4o-mini ~\citep{openai2024gpt4technicalreport} and embedding them using \texttt{sentence\allowbreak-transformers/\allowbreak all\allowbreak-mpnet\allowbreak-base\allowbreak-v2}~\citep{reimers2019sentencebertsentenceembeddingsusing, song2020mpnetmaskedpermutedpretraining}. We then construct a multiplex user graph where facet-wise cosine similarities select up to five neighbors and are averaged for a combined score. Personalization is achieved on \texttt{Qwen/\allowbreak Qwen2.5\allowbreak -7B\allowbreak -Instruct}~\citep{qwen2} via a soft-prompt and LoRA ~\citep{hu2022lora}  finetuning. At inference, cached soft prompts are used with BM25 ~\citep{robertson2009probabilistic}-retrieved history. Additional details are provided in Appendix~\ref{apd:implementation}.

\subsection{Main Results (RQ1 \& RQ2)}

Table~\ref{tab:persona_results} reports performance on three benchmark categories. 
CoPersona consistently achieves the best results across all metrics and datasets. 
On Books, our method surpasses the strongest baseline (DEP) by +3.3 ROUGE-1, +3.4 METEOR, and +3.1 BLEU, indicating substantially better lexical alignment and semantic fidelity. 
Similar gains are observed on Movies \& TV and CDs \& Vinyl, where CoPersona yields the highest ROUGE, METEOR, BLEU, and BERTScore simultaneously.

Notably, while prior methods such as RAG, PAG, and DPL improve over non-personalized generation, their gains remain limited under sparse histories. 
In contrast, CoPersona’s collaborative persona completion provides complementary facet evidence, leading to more accurate and consistent improvements. 

\paragraph{Generalization to Sparse Settings.}
Results on Video Games, Musical Instruments, and Sports \& Outdoors further confirm the robustness of CoPersona, as these datasets are particularly preprocessed to mimic the cold start settings (Section \ref{sec:exp_setup}). 
Our method achieves the highest ROUGE-1, ROUGE-L, and BLEU across all three categories. 
In Video Games and Sports \& Outdoors, CoPersona delivers particularly strong BLEU gains, suggesting improved phrase-level alignment and stylistic consistency.

Although PAG and DPL occasionally yield competitive METEOR scores, they do not consistently outperform CoPersona across metrics. 
This demonstrates that collaborative facet borrowing generalizes beyond the core domains and remains effective across diverse product categories.

\paragraph{LLM-as-Judge Evaluation}

Table~\ref{tab:llm_judge_results} presents human-likeness scores evaluated by Qwen2.5-72B-Instruct-AWQ \cite{qwen2.5,qwen2}. 
CoPersona achieves the highest or tied-highest scores across all evaluated dimensions, indicating that its generated reviews appear more natural and authentic. Due to limited context window constraints, LLM-as-Judge evaluation is performed only on the Books dataset\footnote{The extensive profile sizes and average output lengths in other categories (e.g., Movies \& TV) frequently exceed the model's token capacity}.

\begin{table}[t]
\centering
\small
\setlength{\tabcolsep}{4pt}
\caption{LLM-as-Judge evaluation with Qwen-72B-Instruct-AWQ on \textit{Books} dataset. %
}
\vspace{-5pt}
\label{tab:llm_judge_results}
\begin{tabular}{lccc}
\toprule

\multirow{2}{*}{\textbf{Method}} &
\multicolumn{3}{c}{\textbf{Metrics}} \\
\cmidrule(lr){2-4} 
 & Authenticity& Practical Details& Human Likeness \\
\midrule

 Non-Perso
& 7.70 $\pm$ 0.47& 5.86 $\pm$ 0.68& 6.81$\pm$0.63 \\

 RAG
& 7.87$\pm$0.34& 6.35$\pm$0.81& 7.05$\pm$0.59 \\

 PAG
& 7.89$\pm$0.34& 6.56$\pm$0.87& 7.00$\pm$0.53 \\

 DPL
& 7.86$\pm$0.35& 6.66$\pm$1.12& 7.00$\pm$0.57 \\

 DEP
& 7.87$\pm$0.61& 7.15$\pm$1.39& 7.50 $\pm$0.81\\

 CoPersona (Ours)
& \textbf{7.93$\pm$0.36}& \textbf{8.39$\pm$0.91}& \textbf{7.69$\pm$0.54} \\

\bottomrule
\end{tabular}
\end{table}

Quantitatively, CoPersona consistently outperforms the strongest baseline, DEP, with a notable gain of $+1.24$ in Practical Details. These results demonstrate that collaborative facet completion significantly enhances writing realism and perceived authenticity, confirming that the model's improvements extend beyond mere n-gram overlap.

\subsection{Facet Coverage Bias Analysis }
\label{sec:facet_coverage_analysis}
We numerically evaluate facet coverage bias in the existing dataset. Table~\ref{tab:test_stats} reports the per-domain statistics and the fraction of \emph{facet-biased} samples, defined as users for which at least one facet has support no stronger than \emph{moderate} (Sec.~\ref{sec:facet_discovery}). A higher fraction therefore indicates more severe facet coverage bias (i.e., weaker facet coverage in user histories). We find that this bias is pervasive across domains: {Movies \& TV} exhibits the highest bias rate (100\%), followed by {Books} (97.48\%), while {CDs \& Vinyl} shows a comparatively lower—though still substantial—rate (88.26\%). This suggests domain-dependent evidence sparsity: in {Movies \& TV} and {Books}, nearly all users have at least one facet that is not strongly supported, making the induced profiles more prone to bias from incomplete, diffuse, or noisy support signals.
\begin{table}[H]
\vspace{-5pt}
\centering
\small
\caption{Statistics of the experimented datasets.}
\vspace{-5pt}
\resizebox{0.98\linewidth}{!}{%
\begin{tabular}{lrrrr}
\toprule
\textbf{Category} & \textbf{\#data} & \textbf{Profile Size} & \textbf{Output Length} & \textbf{Facet-biased} \\
\midrule
\textbf{Books}        & 317  & $34.84\pm22.55$ & $1194.90\pm802.44$  & 309 (97.48\%) \\
\textbf{Movies \& TV} & 1925 & $41.11\pm35.90$ & $1704.61\pm1752.44$ & 1925 (100.00\%) \\
\textbf{CDs \& Vinyl} & 1754 & $38.50\pm32.37$ & $1600.04\pm1419.89$ & 1548 (88.26\%) \\
\bottomrule
\end{tabular}%
}
\label{tab:test_stats}
\end{table}

\subsection{Ablation Studies (RQ3)}
We ablate key components of \ours to quantify where its gains come from. Unless otherwise stated, we report results on \textsc{Books} and keep the training setup and decoding strategy fixed.

\paragraph{Dual-branch inference.}
Table~\ref{tab:ablation_books_delta} shows that both branches contribute, with the \emph{parametric} graph branch playing a larger role. Removing the non-parametric branch drops performance from 0.3953 to 0.3822 in ROUGE-1 and from 15.47 to 13.30 in BLEU, indicating that facet-aligned neighbor exemplars provide useful, human-readable evidence beyond the user’s own sparse history.
Removing the parametric branch causes a substantially larger degradation (ROUGE-1 0.3673; BLEU 11.34), suggesting that latent graph reasoning is crucial for denoising and completing weakly supported facets under facet coverage bias. Overall, the best performance is achieved when both branches are combined, validating the complementarity between \emph{text-space} neighbor evidence and \emph{latent-space} collaborative refinement. %

\paragraph{Retrieval strategy for self-history.}
We further examine the impact of the \emph{self-history} retrieval module by replacing the default BM25~\citep{robertson2009probabilistic} retriever with a dense retriever, Contriever~\citep{izacard2021unsupervised}. This substitution leads to a small but consistent decrease in performance (e.g., ROUGE-1 0.3906 vs.\ 0.3953), suggesting that the gains of \ours are not driven by a particular retriever and that the collaborative signal remains effective under alternative retrieval backends. Interestingly, a simple recency-based heuristic attains the best results on \textsc{Books}, likely because recent interactions better capture a user’s current writing style and preferences. Nevertheless, to enable fair comparison with prior work that uses BM25 as the default retriever (e.g., \citet{dpl}), we adopt BM25 as the main system configuration throughout our experiments.

\begin{table}[H]
\centering
\small
\setlength{\tabcolsep}{4pt}
\renewcommand{\arraystretch}{1.15}
\caption{Ablation on Books. Parentheses show absolute changes vs. the full two-branch model.}
\vspace{-6pt}
\resizebox{\columnwidth}{!}{%
\begin{tabular}{lcccc}
\toprule
\textbf{Method} & \textbf{R-1} & \textbf{R-L} & \textbf{MET} & \textbf{BLEU} \\
\midrule
\textbf{Full model} & \underline{0.3953} & \underline{0.2254} & \underline{0.3349} & \underline{15.4703} \\
w/o non-parametric branch
& 0.3822 {\scriptsize(-0.0131)}
& 0.2190 {\scriptsize(-0.0064)}
& 0.3149 {\scriptsize(-0.0200)}
& 13.3008 {\scriptsize(-2.1695)} \\
w/o parametric branch
& 0.3673 {\scriptsize(-0.0280)}
& 0.2067 {\scriptsize(-0.0187)}
& 0.2890 {\scriptsize(-0.0459)}
& 11.3355 {\scriptsize(-4.1348)} \\
w/ Contriever retrieval
& 0.3906 {\scriptsize(-0.0047)}
& 0.2226 {\scriptsize(-0.0028)}
& 0.3337 {\scriptsize(-0.0012)}
& 15.2739 {\scriptsize(-0.1964)} \\
w/ Recency retrieval
& \textbf{0.4062} {\scriptsize(+0.0109)}
& \textbf{0.2408} {\scriptsize(+0.0154)}
& \textbf{0.3520} {\scriptsize(+0.0171)}
& \textbf{16.6532} {\scriptsize(+1.1829)} \\
\bottomrule
\end{tabular}%
}
\label{tab:ablation_books_delta}
\end{table}

\begin{figure}[t]
  \centering
  \begin{subfigure}[t]{0.32\linewidth}
    \centering
    \includegraphics[width=\linewidth]{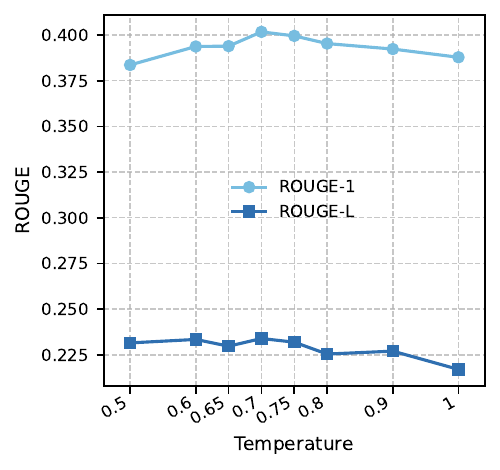}
  \end{subfigure}\hfill
  \begin{subfigure}[t]{0.32\linewidth}
    \centering
    \includegraphics[width=\linewidth]{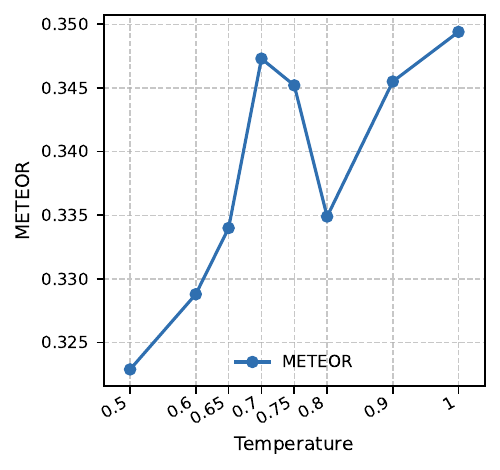}
  \end{subfigure}\hfill
  \begin{subfigure}[t]{0.32\linewidth}
    \centering
    \includegraphics[width=\linewidth]{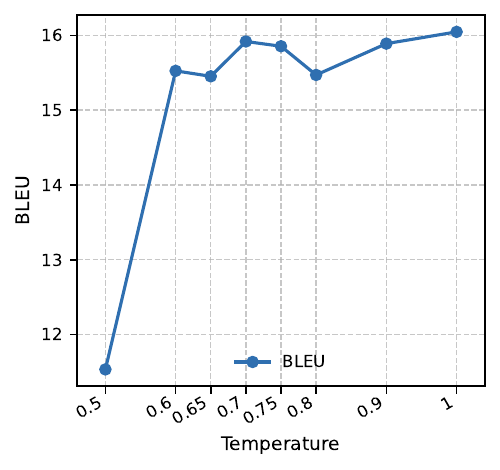}
  \end{subfigure}
  \vspace{-10pt}
  \caption{Sensitivity to decoding temperature. }
  \label{fig:temp-sensitivity}
    \vspace{-10pt}
\end{figure}

  \vspace{-10pt}
  
\begin{figure}[t]
    \centering
    \includegraphics[width=0.32\linewidth]{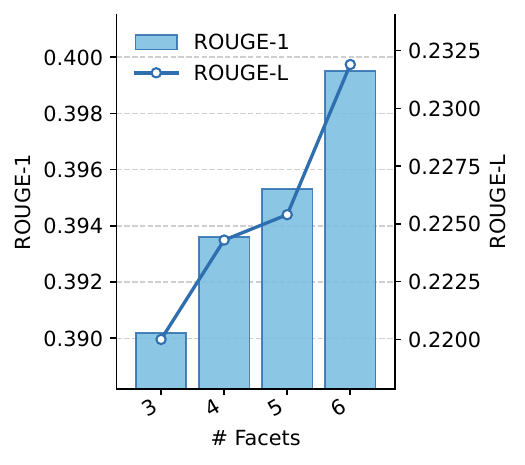}\hfill
    \includegraphics[width=0.32\linewidth]{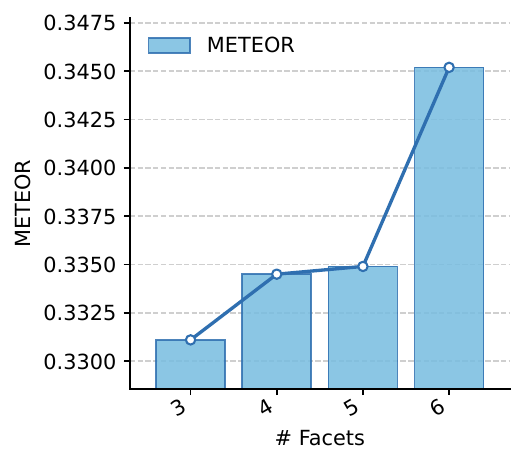}\hfill
    \includegraphics[width=0.32\linewidth]{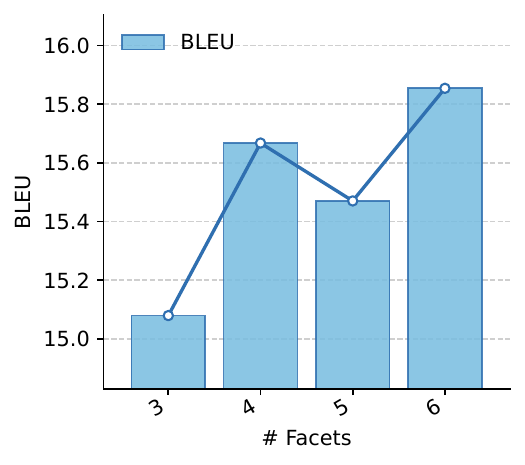}
    \vspace{-10pt}
    \caption{Ablation on the number of induced facets (Books).}
    \label{fig:facet_ablation}
    \vspace{-5pt}
\end{figure}

  \vspace{-10pt}
  
\begin{figure}[t]
    \centering
    \includegraphics[width=0.32\linewidth]{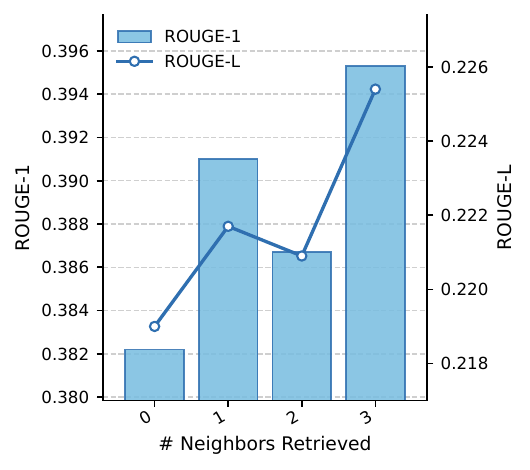}\hfill
    \includegraphics[width=0.32\linewidth]{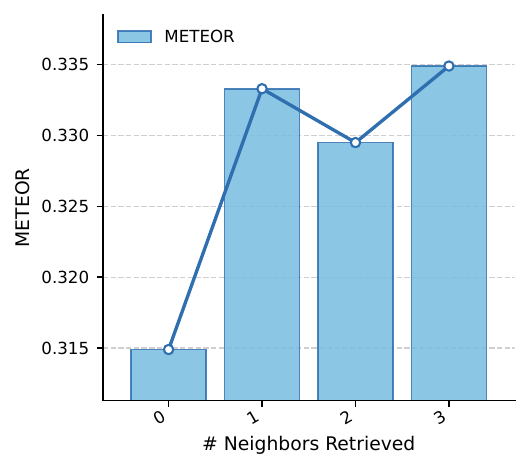}\hfill
    \includegraphics[width=0.32\linewidth]{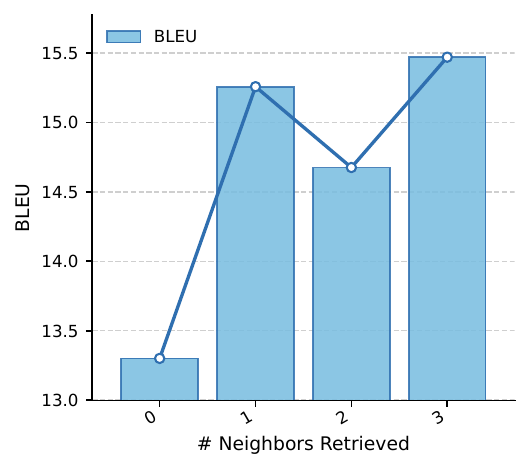}
    \vspace{-10pt}
    \caption{Ablation on the number of retrieved neighbors (Books).}
    \label{fig:neighbor_count_ablation}
      \vspace{-10pt}
\end{figure}

\subsection{Hyperparameter Sensitivity (RQ4)}
We examine robustness to several key design choices on \textsc{Books}.
\paragraph{Decoding temperature.}
Figure~\ref{fig:temp-sensitivity} varies the decoding temperature over a broad range (0.5--1.0). Performance is largely stable in the range 0.7--1.0, with a mild peak at moderate temperatures, indicating that \ours does not require careful tuning of sampling stochasticity to maintain strong quality. Following prior work~\citep{dep}, we use a default temperature of 0.8 in all experiments.

\paragraph{Number of induced facets.}
Figure~\ref{fig:facet_ablation} varies the size of the induced facet schema (3--6). We observe a gradual improvement as the number of facets increases, consistent with the intuition that a more fine-grained schema better decomposes a user’s persona and reduces facet mixing when evidence is uneven. At the same time, gains diminish beyond a moderate number of facets, suggesting that \ours benefits from finer structure but does not rely on an aggressively large schema.

\paragraph{Retrieved neighbor budget.}
Figure~\ref{fig:neighbor_count_ablation} varies the number of retrieved neighbors (0--3). Performance improves steadily as we retrieve more neighbors, indicating that additional peer evidence provides complementary signals that help complete under-supported facets. The improvement saturates at larger budgets, implying that only a small but related neighborhood is sufficient to capture most of the collaborative benefit while keeping inference efficient.

\subsection{Discovered Facet Schema}
To better understand the personas induced by \ours, we present the facets discovered across domains (Table~\ref{tab:facet_discovery}). The induced schemas vary by dataset and reflect domain-specific user concerns: in narrative domains such as Books and Movies, facets often capture emotional engagement and critical interpretation, whereas in product-centric domains such as Sports \& Outdoors and Musical Instruments, they emphasize utility-oriented considerations (e.g., value for money, product quality, and usage context). Overall, these patterns suggest that our facet discovery adapts to domain signals and yields interpretable preference dimensions for personalization, rather than dataset-agnostic latent factors.

\begin{table*}[t]
\centering
\small
\setlength{\tabcolsep}{5pt}
\caption{
Persona facets discovered across different domains.
}
\begin{tabular}{p{3cm} p{12.5cm}}
\toprule
\textbf{Dataset} & \textbf{Discovered Facets} \\
\midrule

\textbf{Books}
& Engagement Level; Genre Preference; Critical Tone; Review Length; Diversity of Themes. \\

\textbf{Movies \& TV}
& Emotional Engagement; Critical Analysis; Genre Preference; Rating Consistency; Social Influence. \\

\textbf{CDs \& Vinyl}
& Rating Enthusiasm; Critical Engagement; Nostalgia Factor; Genre Preference; Review Style. \\

\textbf{Musical Instruments}
& Satisfaction Level; Product Knowledge; Value for Money; Usage Context; Brand Loyalty. \\

\textbf{Sports \& Outdoors}
& Product Satisfaction; Quality Perception; User Engagement; Value for Money; Brand Loyalty. \\

\textbf{Video Games}
& Engagement Level; Critical Perspective; Nostalgia Factor; Community Engagement; Gameplay Preference. \\

\bottomrule
\end{tabular}

\label{tab:facet_discovery}
\end{table*}

\section{Related Work}

\paragraph{Personalization in LLMs.} LLM personalization seeks to adapt model behavior to individual users’ preferences, goals, and interaction patterns~\citep{lamp}. Prior work generally falls into two broad paradigms: prompting-based methods and finetuning-based techniques. Early prompting-based approaches~\citep{hwang2023aligning, kang2023llms} encode user information directly in the input, using structured or unstructured natural language prompts to summarize preferences or interaction history, but they rely heavily on careful prompt construction and aggressive compression of user data~\citep{li2024learning, mao2025reinforced}. More recent work~\citep{lamp, salemi2024optimization, yang2023palr, li2023gpt4rec, huang2023learning, zerhoudi2024personarag, wang2024unims, longlamp} adopts retrieval-augmented strategies that dynamically incorporate relevant signals from a user’s past interactions into the model context at inference time, alleviating context-length constraints while remaining vulnerable to retrieval errors, particularly when user histories are sparse or noisy. Finetuning-based approaches, including soft-prompt~\citep{hebert2024persoma, liu2025llms+}, LoRA~\citep{tan2024democratizing, zhuang2024hydra, tan2024personalized, zhang2024personalized}, and RLHF-style~\citep{stiennon2020learning, jang2023personalized, park2024rlhf, poddar2024personalizing, li2024personalized, zhu2024personality} finetuning, offer an alternative by embedding user-specific information directly into model parameters, enabling more efficient and persistent personalization. However, these works are typically limited to capturing single user's information.

\paragraph{Cold-Start Problem and Collaborative Filtering.} The cold-start problem, long studied in recommendation systems, arises when limited interaction data prevents accurate preference modeling for new users~\citep{lika2014facing, schein2002methods}. Common solutions include leveraging side information~\citep{abdullah2021eliciting}, and collaborative filtering~\citep{bobadilla2012collaborative}, which mitigates sparsity by transferring signals from similar users. Inspired by this paradigm, recent work extends collaborative filtering to LLM personalization by incorporating inter-user information. One line of work explicitly models user–peer differences, using natural-language summaries or difference-aware latent soft prompts~\citep{dpl, dep}. Other approaches go further by directly injecting peer interactions into prompts or constructing shared persona representations~\citep{shi2025retrieval, sun2025persona, yazan2025improving}. However, they typically rely on global similarity in an unstructured space and perform only shallow retrieval over peers’ histories~\citep{shi2025retrieval}. In contrast, our approach adopts a structured, facet-aware framework: we construct a multiplex facet-level graph and borrow signals from neighbors aligned on specific facets via both non-parametric retrieval and parametric message passing, enabling targeted transfer rather than indiscriminate peer aggregation.

\begin{acks}
This work was supported in part by the 2025 Samsung Leap-U Program.
\end{acks}

\bibliographystyle{ACM-Reference-Format}
\bibliography{ref}

\clearpage
\appendix
\section{Overview of Datasets}
\label{apd_dataset}

We study personalized review generation using Amazon Reviews 2023~\cite{amazon2023} following the DPL preprocessing protocol~\cite{dpl}. 
Besides the original DPL splits, we construct three additional categories (\textit{Video Games}, \textit{Sports \& Outdoors}, \textit{Musical Instruments}) with adjusted thresholds to better simulate sparse-user settings.

\paragraph{Preprocessing}

We filter reviews to keep only entries with valid text, title, rating, and timestamp above a minimum length. 
Items are kept if their cleaned descriptions fall within a length range and have valid metadata.
We then iteratively prune the user--item graph so that:
(i) each item has at least a minimum number of reviewers, and  
(ii) each user has a minimum number of reviews.  
Duplicate reviews are removed by keeping the most recent unique (user, item) interaction.

\paragraph{Profile Construction}

For each user, reviews are sorted chronologically. 
The most recent $k=2$ interactions are held out as targets (bounded by at most half of the history). 
Earlier interactions form the \emph{profile}. 
Users with too few profile reviews and items appearing in too few profiles are further pruned.

\paragraph{Splits}

Given a profile and held-out interactions:
\begin{itemize}[leftmargin=*, itemsep=2pt]
    \item \textbf{Train:} predict earlier held-out reviews autoregressively,
    \item \textbf{Validation:} predict the second-to-last review,
    \item \textbf{Test:} predict the last review.
\end{itemize}

\begin{table}[h]
\centering
\small
\caption{Main preprocessing thresholds.}
\label{dataset_stat}
\begin{tabular}{lccc}
\toprule
Parameter & VG & Sports & Music \\
\midrule
Min review length & 110 & 90 & 90 \\
Min desc length & 60 & 50 & 50 \\
Min users/item & 4 & 3 & 3 \\
Min reviews/user & 8 & 5 & 5 \\
Holdout $k$ & 4 & 3 & 3 \\
Min profile size & 6 & 4 & 4 \\
\bottomrule
\end{tabular}
\end{table}

\begin{table}[H]
\centering
\small
\caption{Statistics of the experimented datasets.}
\vspace{-5pt}
\resizebox{0.98\linewidth}{!}{%
\begin{tabular}{lrrr}
\toprule
\textbf{Category} & \textbf{\#data} & \textbf{Profile Size} & \textbf{Output Length} \\
\midrule
\textbf{Video\_Games}        & 837& 15.22$\pm$10.90& 1368.57$\pm$1861.22\\
\textbf{Musical Instruments}& 556& $11.94\pm6.43$& $716.19\pm706.38$\\
\textbf{Sports\_and\_Outdoors}& 531& 18.18$\pm$16.69& 578.59$\pm$443.18\\
\bottomrule
\end{tabular}%
}
\label{tab:test_stats_new}
\end{table}

\paragraph{Data Format}

Each instance contains a user ID, profile reviews, and a target review:

\begin{quote}
\small
\texttt{\{user\_id, profile: [reviews...], data: target\_review\}}
\end{quote}

Each input concatenates target item title, description, review title, rating, and retrieved user history.

\section{Baseline Details}\label{apd_baseline}

We compare \ours with representative personalization baselines that differ in how user information is represented and incorporated into generation.

\begin{itemize}[leftmargin=*]

\item \textbf{Non-Perso}.
This baseline removes personalization signals and conditions the LLM only on item-level inputs, the target rating, and the review title. It serves as a lower bound for generation without user history.

\item \textbf{RAG}~\cite{lamp}.
RAG retrieves a small set of the user's recent reviews and concatenates them into the prompt, allowing the LLM to infer preferences directly from raw historical examples.

\item \textbf{PAG}~\cite{pag}.
PAG summarizes recent reviews into a natural-language user profile and provides it together with retrieved reviews, combining high-level preferences with concrete evidence.

\item \textbf{DPL}~\cite{dpl}.
DPL constructs a difference-aware profile by comparing the target user with representative users selected through embedding-space clustering. The resulting summary captures differences in aspects such as sentiment, topics, and style, and is prepended to the generation prompt. For fairness, we use the same embedding model as in \ours for representative-user selection.

\item \textbf{DEP}~\cite{dep}.
DEP learns latent user embeddings that capture preference deviations from other users and injects the resulting personalization vectors into a frozen LLM as soft prompts, enabling representation-level personalization without increasing prompt length.

\end{itemize}

\section{Implementation Details}
\label{apd:implementation}
\paragraph{Facet Disentanglement.}
We implement a two-stage pipeline for persona facet discovery.
First, we construct a user-level document by concatenating up to 8 recent reviews per user (including rating, title, and text) and truncating to a fixed length for stability.
We encode each user document using a SentenceTransformer encoder (\textit{all-MiniLM-L6-v2}) with $\ell_2$-normalized embeddings~\citep{reimers2019sentencebertsentenceembeddingsusing}.
We augment semantic embeddings with lightweight writing-style features (e.g., average review length, sentence-length proxy, punctuation/emoji rate, lexical diversity, and rating statistics), which are standardized and concatenated to the embedding vector.
We 
then cluster users using KMeans~\citep{Hartigan1979KMeans}, and
for each cluster, we compute representative keywords via TF--IDF~\citep{jm3} and select representative users via medoid-based selection using cosine distance.
Finally, we pack cluster summaries (cluster size, top terms, representative excerpts, and style statistics) and prompt an LLM to induce a small, non-redundant set of human-interpretable facets, each with a definition, value specification, scoring rubric, and evidence grounded in specific clusters; the model returns a structured facet schema in JSON for downstream personalization.
For efficient local inference, we support vLLM~\citep{vllm} as the backend.

\paragraph{Facet summarization and per-facet embeddings.}
We start from a discovered facet set. For each user, we summarize up to 6 historical reviews into facet-specific profile text using \texttt{gpt-4o-mini}~\citep{openai2024gpt4technicalreport}, with temperature 0.2 and a 2048-token budget, batching 8 users per LLM call. We then embed each facet summary with \texttt{sentence\allowbreak-transformers/\allowbreak all\allowbreak-mpnet\allowbreak-base\allowbreak-v2}~\citep{reimers2019sentencebertsentenceembeddingsusing, song2020mpnetmaskedpermutedpretraining}.\footnote{\url{https://huggingface.co/sentence-transformers/all-mpnet-base-v2}} Embeddings are computed with max sequence length 512, batch size 32, and L2 normalization. Each per-user facet embedding concatenates the facet paragraph with up to two evidence quotes when available.

\paragraph{Multiplex user--user similarity graph.}
We construct a multiplex graph where each facet induces a similarity layer. For each facet, we compute cosine similarity (dot product of normalized embeddings), retain up to 5 neighbors per facet with similarity $\ge 0.35$, and store the facet-specific similarity as an edge weight. A user pair’s combined similarity is defined as the mean of its facet weights. We use fixed reliability weights $\boldsymbol{\omega}=[0.05,\,0.25,\,0.60,\,0.95]$ in graph construction stage, corresponding to $r\in[\texttt{none},\,\texttt{weak},\,\texttt{moderate},\,\texttt{strong}]$, respectively.

\paragraph{Monotone label-to-gate mapping. }
In learnable parametric branch, we convert the ordinal reliability label $r_u^{(m)} \in\{$ none, weak, moderate, strong $\}$ into a scalar trust gate $\gamma_u^{(m)} \in[0,1]$ via a lightweight learnable mapping. Let $\ell(r) \in\{0,1,2,3\}$ denote the ordinal encoding of the label. We learn a gate value $\left\{\gamma_k\right\}_{k=0}^3$ for each label level and set $\gamma_u^{(m)}=\gamma_{\ell\left(r_u^{(m)}\right)}$. To ensure $\gamma_k$ is monotonically non-decreasing with evidence strength, we parameterize it using cumulative non-negative increments:
$
\delta_i=\operatorname{softplus}\left(\tilde{\delta}_i\right) \geq 0, \quad i=1,2,3, \quad z_k=z_0+\sum_{i=1}^k \delta_i, \quad \gamma_k=\sigma\left(z_k\right)
$
where $z_0 \in \mathbb{R}$ and $\tilde{\delta}_i \in \mathbb{R}$ are learnable parameters, $\sigma(\cdot)$ is the sigmoid function, and softplus  enforces $\delta_i \geq 0$. This guarantees $\gamma_0 \leq \gamma_1 \leq \gamma_2 \leq \gamma_3$ while keeping the mapping differentiable and adding only four scalar parameters (shared across facets in our implementation; a per-facet variant is straightforward).

\paragraph{Personalizer fine-tuning.}
We use \texttt{Qwen/\allowbreak Qwen2.5\allowbreak -7B\allowbreak -Instruct}~\citep{qwen2, qwen2.5} as the base model. Our soft-prompt personalizer is a two-layer MLP that maps each user’s facet representation to a sequence of virtual tokens. We first construct a user vector using a single-layer reliability-gated graph encoder (hidden size 512, ReLU, dropout 0.1). The projector uses a two-layer MLP:
\texttt{Linear(d\_in -> 512)} + \texttt{Tanh} + \texttt{Linear(512 -> L*H)},
where $d_{\mathrm{in}}$ is the user-vector dimension, $L{=}20$ is the soft-prompt length, and $H$ is the LLM hidden size (Qwen2.5-7B). The output is reshaped to $L\times H$ and prepended to the token embeddings. 

During training, we prepend \texttt{soft\_prompt} to the token embeddings, extend the attention mask with ones for the virtual tokens, and set the corresponding labels to $-100$ so that the loss is computed only over real tokens. We fine-tune the base model with LoRA~\citep{hu2022lora} (rank 16, $\alpha{=}32$, dropout 0.05), while optimizing the soft-prompt personalizer and graph encoder end-to-end. At inference time, we generate the soft prompt once per user and cache it for reuse.

We train for 10 epochs with batch size 1 and gradient accumulation of 4 (effective batch size 4), using AdamW~\citep{loshchilov2017decoupled} with learning rate $2\times10^{-5}$, weight decay 0.01, and betas $(0.9, 0.98)$. We use a cosine learning-rate schedule with warmup ratio 0.06 and clip gradients to a maximum norm of 1.0. At inference, prompts include four retrieved reviews and up to three neighbors selected by the highest combined similarity. To fit within the model’s context window, we cap the input at 8192 tokens and apply per-component limits: 1024 tokens per retrieved review, 1024 tokens for the user’s historical reviews, and 512 tokens per neighbor facet summary.

\section{Computational Cost Analysis}
Most of the computation in our personalization pipeline can be performed offline and cached. Facet summarization, embedding extraction, and graph construction are per-user (or per-refresh) preprocessing steps, and the graph message passing can likewise be run ahead of time to produce a fixed user representation. We then precompute the user’s soft prompt (virtual tokens) once and store it. As a result, online inference only needs to retrieve the cached soft prompt and prepend it to the decoding context, without running any graph/message-passing or embedding-related computation on the critical path; consequently, the incremental runtime overhead during generation is negligible in practical deployments.

\end{document}